\documentclass[runningheads]{llncs}
\usepackage{graphicx}
\usepackage[dvipsnames]{xcolor}

\usepackage{amssymb,amsmath}

\usepackage{tikz}
\usetikzlibrary{calc}

\begin{document}
\title{Quantum advantage by relational queries about physically realizable equivalence classes}
%
%
\author{Karl Svozil\orcidID{0000-0001-6554-2802}}
\authorrunning{Karl Svozil}
%
\institute{Institute for Theoretical Physics,
Vienna  University of Technology,
Wiedner Hauptstrasse 8-10/136,
1040 Vienna,  Austria  \\
\email{svozil@tuwien.ac.at}\\
\url{http://tph.tuwien.ac.at/\textasciitilde{}svozil}
}
\maketitle              
\begin{abstract}
Relational quantum queries are sometimes capable to effectively decide between collections of mutually exclusive elementary cases without completely resolving and determining those individual instances. Thereby the set of mutually exclusive elementary cases is effectively partitioned into equivalence classes pertinent to the respective query. In the second part of the paper, we review recent progress in theoretical certifications (relative to the assumptions made) of quantum value indeterminacy as a means to build quantum oracles for randomness.

\end{abstract}

\section{Quantum (dis-)advantages}

Contemporary quantum information theory appears to be challenging yet far from being fully comprehended, worked out and mature.
It is based on quantum mechanics, a theory whose semantics has been notoriously debated almost from its inception, while its syntax -- its formalism, and, in particular,
the rules of deriving predictions
-- are highly successful, accepted and relied upon.
Depending on temperament and metaphysical inclination, its proponents admit that nobody understands quantum mechanics~\cite{feynman-law,clauser-talkvie},
maintain that there is no issue whatsoever~\cite{Englert2013,fuchs-peres},
one should not bother too much~\cite{dirac,bell-a} about its meaning and foundations, and rather shut up and calculate~\cite{mermin-1989-shutup,mermin-2004-shutup}.

By transitivity or rather reduction, quantum information theory inherits quantum mechanics'
apparent lack of consensus, as well as a certain degree of cognitive dissonance between applying the formalism while suffering from an absence of conceptual clarity~\cite{mermin-2019},
Strong hopes, claims and promises~\cite{svozil-2016-quantum-hokus-pokus} of quantum supremacy are
accompanied by the pertinent question of what exactly, if at all, could make quantum information and computation outperform classical physical resources.
Surely many nonclassical quantum features present themselves as being useful or decisive in this respect; among them complementarity, coherence (aka parallelism), entanglement,
or value indeterminacy (aka contextuality).
But if and how exactly those features will contribute or enable future algorithmic advances still remains to be seen.

The situation is aggravated by the fact that, although the quantum formalism amounts to linear algebra and functional analysis,
some of its most important theorems are merely superficially absorbed by the community at large: take,
for example, Gleason's theorem~\cite{Gleason}, and extensions thereof~\cite{pitowsky:218,2015-AnalyticKS}.
Another example is Shor's factoring algorithm~\cite[Chapter~5]{nielsen-book10} whose presentations often suffer from the fact
that its full comprehension requires a nonsuperficial understanding of number theory,
analysis, as well as quantum mechanics; a condition seldom encountered in a single (wo)man.
Moreover, often one is confronted with confusing opinions: for instance, the claim that quantum computation is universal with respect to either unitary transformations
or first-order predicate calculus is sometimes confused with full Turing universality.
And the plethora of algorithms collected into a quantum algorithm zoo~\cite{jordan-zoo} is compounded by the quest of exactly why and how quantum algorithms may outperform classical ones.

It also may very well be that the necessity to maintain coherence throughout a quantum computation will impose an exponential overhead of ``physical stuff'' necessary to achieve this goal.
This could well compensate or even outweigh quantum parallelism; that is, the exponential simultaneous
co-representability of (coherent superpositions of) classical mutually exclusive cases of a computation.
Nevertheless, in what follows we shall consider the feasibility
for speedups from such quantum computational strategies involving quantum parallelism.

\section{Suitable partitioning of cases}

One quantum feature called ``quantum parallelism,''
which is often presented as a possible quantum resource not available classically,
is the capacity of $n$ quantum bits to encode $2^n$ classically
mutually exclusive distinct classical bit states at once, that is, simultaneously:
$
\vert \Psi \rangle = \sum_{i=0}^{2^n-1} \psi_i \vert i \rangle
$,
where the index $i$ runs through all $2^n$ possible combinations of $n$ classically mutually exclusive bit states $\{0,1\}$,
$\vert i \rangle $ are elements of an orthonormal basis in $2^n$-dimensional Hilbert space,
and $\psi_i$ represent probability amplitudes whose absolute squares sum up to $1$.
This seemingly absurd co-representability of contradicting classical states was the motivation for Schr\"odinger's cat paradox~\cite{schrodinger}.

The state
$
\vert \Psi \rangle
$
``carrying'' all these classical cases in parallel is not directly accessible or ``readable'' by any physical operational means.
And yet, it can be argued that its simultaneous representation of classically exclusive cases can be put to practical use indirectly if certain criteria are met:
\begin{itemize}
\item
first of all, there needs to be a quantum physical realizable grouping or partitioning of the
classical cases, associated with a particular query of interest; and
\item
second, this aforementioned query needs to be realizable by a quantum observable.
\end{itemize}
In that way, one may attain knowledge of a particular feature one is interested in; but,
unlike classical computation, (all) other features remain totally unspecified and unknown.
There is no ``free quantum lunch'' here, as a total specification of all observables would
require the same amount of quantum queries as with classical resources.
And yet, through coherent superposition (aka interference)
one might be able to ``scramble'' or re-encode the signal
in such a way that some features can be read off of it very efficiently --
indeed, with an exponential (in the number of bits) advantage over classical computations
which lack this form of rescrambling and re-encoding (through coherent superpositions).
However, it remains to be seen whether, say, classical analog computation with waveforms,
can produce similar advantages.

For the sake of a demonstration, the Deutsch algorithm~\cite[Chapter~2]{mermin-07} serves as a Rosetta stone of sorts
for a better understanding of the formalism and respective machinery at work in such cases.
It is based on the four possible binary functions $f_0, \ldots , f_3$ of a single bit $x \in \{0,1\}$:
the two constant functions $f_0(x)=1-f_3(x)=0$,
as well as the two nonconstant functions: the identity $f_1(x)=x$ and the not $f_3(x) = (x+1) \text{ mod } 2$, respectively.
Suppose that one is presented with a black box including in- and output interfaces,
realizing one of these classical functional cases, but it is unknown which one.
Suppose further that one is only interested in their parity; that is, whether or not the encoded black box function is a constant function of the argument.
Thereby, with respect to the corresponding equivalence relation of being ``(not) constant in the arguement''
the set of functions
$\{f_0, \ldots , f_3\}$
is partitioned into
$\{\{f_0,f_3\},\{f_1,f_2\}\}$.

A different way of looking at this relational encoding
is in terms of zero knowledge proofs: thereby
nature is in the role of an agent
which is queried about a property/proposition, and issues a correct answer
without disclosing
all the details and the fine structure of the way this result is obtained.

Classically the only way of figuring this (``constant or not'')   out
is to input the two bit-state cases, corresponding to two separate queries.
If the black box admits quantum states,
then the Deutsch algorithm presents a way to obtain the answer (``constant or not'')
directly in one query.
In order to do this one has to perform
three successive steps~\cite{svozil-2005-ko,2007-tkadlec-svozil-springer}:
\begin{itemize}
\item
first one needs to scramble the classical bits into a coherent superposition of the two classical bit states.
This can be done by a Hadamard transformation, or a quantum Fourier transformation;
\item
second, one has to transform the coherent superposition according to the binary function which is encoded in the box.
This has to be done while maintaining reversibility; that is, by taking ``enough'' auxiliary bits
to maintain bijectivity/permutation; even if the encoded function is many-to-one (eg, constant).
\item
third, one needs to unscramble this resulting state to produce a classical output signal which indicates the result of the query.
As all involved transformations need to be unitary and thus reversible
the latter task can again be achieved by an (inverse) Hadamard transformation, or an (inverse)
quantum Fourier transformation.
\end{itemize}

This structural pattern repeats itself in many quantum algorithms suggested so far.
It can be subsumed into the threefold framework:
``spread (the classical state into a coherent superposition of classical states)
---
transform (according to some functional form pertinent to the problem or query considered)
---
fold (into partitions of classical states which can be accessed via quantum queries and yield classical signals).''

Besides the (classical) pre- and post-processing of the data,
Shor's algorithm~\cite[Chapter~5]{nielsen-book10}
has a very similar structure in its quantum (order-finding) core:
It creates a superposition of classically mutually exclusive states $i$
{\it via} a generalized Hadamard transformation.
It then processes this coherent superposition of all $i$ by computing $x^i \text{ mod } n$,
for some (externally given) $x$ and $n$, the number to be factored.
And it finally ``folds back'' the expanded, processed state by applying an inverse quantum Fourier transform,
which then (with high probability) conveniently yields a classical information (in one register) about the period or order; that is,
the least positive integer $k$ such that $x^k =1 (\text{mod } n)$ holds.
As far as Shor's factoring algorithm is concerned, everything else is computed classically.

Whether or not this strategy to find ``quantum oracles'' corresponding to arbitrary partitions
of classical cases is quantum feasible remains to be seen. There appears to be an {\it ad hoc} counterexample,
as there is no speedup for generalized parity~\cite{Farhi-98}; at least with the means considered.

\section{Quantum oracles for random numbers}

\begin{figure}[h]
\begin{center}
\begin{tikzpicture}  [scale=0.3]
\labeltrue
        \tikzstyle{every path}=[line width=1.5pt]
        \tikzstyle{c1}=[circle,fill,inner sep=3]
        \tikzstyle{c2}=[circle,fill,inner sep=2]
        \tikzstyle{c3}=[circle,fill,inner sep=1]
        \tikzstyle{s1}=[color=red,rectangle,minimum size=8,inner sep=6]
        \tikzstyle{d1}=[draw=none,circle,minimum size=4]
        \tikzstyle{e1}=[color=gray,rectangle,minimum size=8,inner sep=6]


\draw [color=orange]  (4,0)  coordinate[c1,fill,label=225:{\color{black}  $\vert {\bf b} \rangle $}] (b) -- (13,0)    coordinate[c1,fill,label={[label distance=-1]270:{\iflabel \footnotesize \color{black}  $2$\fi}}] (2) -- (22,0)  coordinate[c1,fill,label={[label distance=-1]315:{\iflabel \footnotesize \color{black}  $3$\fi}}] (3);
\draw [color=blue] (3) -- (26,12)  coordinate[c1,fill,pos=0.8,label={[label distance=-1]0:{\iflabel \footnotesize \color{black}  ${21}$\fi}}] (21) coordinate[c1,fill,label={[label distance=-3]0:{\iflabel \footnotesize \color{black}  ${23}$\fi}}] (23);
\draw [color=green] (23) -- (22,18.5) coordinate[c1,fill,pos=0.4,label={[label distance=-1]0:{\iflabel \footnotesize \color{black}  ${29}$\fi}}] (29) coordinate[c1,fill,label={[label distance=-1]45:{\iflabel \footnotesize \color{black}  $5$\fi}}] (5);
\draw [color=magenta] (5)-- (13,18.5)coordinate[c1,fill,label=90:{\color{black}  $\vert {\bf a} \rangle $}] (a) -- (4,18.5)  coordinate[c1,fill,label={[label distance=-1]135:{\iflabel \footnotesize \color{black}  $4$\fi}}] (4);
\draw [color=CadetBlue] (4) -- (0,12)   coordinate[c1,fill,pos=0.6,label={[label distance=-1]180:{\iflabel \footnotesize \color{black}  ${10}$\fi}}] (10)  coordinate[c1,fill,label={[label distance=-1]180:{\iflabel \footnotesize \color{black}  $7$\fi}}] (7);
\draw [color=brown] (7) -- (b)      coordinate[c1,fill,pos=0.2,label={[label distance=-1]180:{\iflabel \footnotesize \color{black}  $6$\fi}}] (6);

        \draw [color=gray] (a) -- (2) coordinate[c1,fill,pos=0.52,label={[label distance=-1, yshift=2]357.5:{\iflabel \footnotesize \color{black}  $1$\fi}}] (1);

        \draw [color=violet] (5) -- (22,6) coordinate[c1,fill,pos=0.4,label={[label distance=-1]0:{\iflabel \footnotesize \color{black}  ${11}$\fi}}] (11) coordinate[c1,fill,label={[label distance=-1]0:{\iflabel \footnotesize \color{black}  $9$\fi}}] (9);

\draw [color=Apricot] (9) -- (b) coordinate[c1,fill,pos=0.3,label={[label distance=-1]280:{\iflabel \footnotesize \color{black}  $8$\fi}}] (8);

\draw [color=TealBlue] (4) -- (4,6) coordinate[c1,fill,pos=0.4,label={[label distance=-1]180:{\iflabel \footnotesize \color{black}  ${28}$\fi}}] (28) coordinate[c1,fill,label={[label distance=-3]180:{\iflabel \footnotesize \color{black}  ${22}$\fi}}] (22);
\draw [color=YellowGreen] (22) -- (3) coordinate[c1,fill,pos=0.2,label={[label distance=-1]260:{\iflabel \footnotesize \color{black}  ${19}$\fi}}] (19);

        \coordinate (25) at ([xshift=-4cm]1);
        \coordinate (27) at ([xshift=4cm]1);

\draw [color=MidnightBlue]  (22) -- (25) coordinate[c1,fill,pos=0.5,label={[label distance=-1]180:{\iflabel \footnotesize \color{black}  ${24}$\fi}}] (24) coordinate[c1,fill,label={[label distance=-1]180:{\iflabel \footnotesize \color{black}  ${25}$\fi}}] (25);
\draw [color=Mulberry] (25) -- (9) coordinate[c1,fill,pos=0.8,label={[label distance=-1]90:{\iflabel \footnotesize \color{black}  ${35}$\fi}}] (35);

\draw [color=BrickRed]  (7) -- (27) coordinate[c1,fill,pos=0.5,label={[label distance=-3]90:{\iflabel \footnotesize \color{black}  ${34}$\fi}}] (34) coordinate[c1,fill,label={[label distance=-1]90:{\iflabel \footnotesize \color{black}  ${27}$\fi}}] (27);
\draw [color=Emerald] (27) -- (23) coordinate[c1,fill,pos=0.25,label={[label distance=-1]320:{\iflabel \footnotesize \color{black}  ${26}$\fi}}] (26);

\draw [color=BlueGreen]  (10) -- (15.5,17.5) coordinate[c1,fill,pos=0.5,label={[label distance=-1]90:{\iflabel \footnotesize \color{black}  ${12}$\fi}}] (12) coordinate[c1,fill,label={[label distance=-1,xshift=5]270:{\iflabel \footnotesize \color{black}  ${13}$\fi}}] (13);
\draw [color=Tan] (13) -- (29) coordinate[c1,fill,pos=0.4,label={[label distance=-1]90:{\iflabel \footnotesize \color{black}  ${31}$\fi}}] (31);

\draw [color=RawSienna]  (28) -- (10.5,15) coordinate[c1,fill,pos=0.5,label={[label distance=-3, yshift=-3]160:{\iflabel \footnotesize \color{black}  ${30}$\fi}}] (30) coordinate[c1,fill,label={[label distance=-5]45:{\iflabel \footnotesize \color{black}  ${15}$\fi}}] (15);
\draw [color=SpringGreen] (15) -- (11) coordinate[c1,fill,pos=0.6,label={[label distance=-1]90:{\iflabel \footnotesize \color{black}  ${14}$\fi}}] (14);

\draw [color=Salmon]  (15) -- (1) coordinate[c1,fill,pos=0.2,label={[label distance=-1, yshift=2]180:{\iflabel \footnotesize \color{black}  ${17}$\fi}}] (17);
\draw [color=Fuchsia] (1)-- (13) coordinate[c1,fill,pos=0.3,label={[label distance=-1]0:{\iflabel \footnotesize \color{black}  ${16}$\fi}}] (16);

\draw [color=CornflowerBlue]  (19) -- (16) coordinate[c1,fill,pos=0.3,label={[label distance=-1]180:{\iflabel \footnotesize \color{black}  ${18}$\fi}}] (18);
\draw [color=pink] (16) -- (8) coordinate[c1,fill,pos=0.7,label={[label distance=-1]180:{\iflabel \footnotesize \color{black}  ${32}$\fi}}] (32);

\draw [color=PineGreen]  (6) -- (17) coordinate[c1,fill,pos=0.7,label={[label distance=-1, yshift=2]180:{\iflabel \footnotesize \color{black}  ${33}$\fi}}] (33);
\draw [color=DarkOrchid] (17) -- (21) coordinate[c1,fill,pos=0.4,label={[label distance=-3]20:{\iflabel \footnotesize \color{black}  ${20}$\fi}}] (20);

\draw [color=black] (25) -- (1) -- (27);

\draw (a) coordinate[c2,fill=gray];

\draw (b) coordinate[c2,fill=brown];
\draw (b) coordinate[c3,fill=Apricot];

\draw (4) coordinate[c2,fill=CadetBlue];
\draw (4) coordinate[c3,fill=TealBlue];

\draw (5) coordinate[c2,fill=magenta];
\draw (5) coordinate[c3,fill=violet];

\draw (7) coordinate[c2,BrickRed];
\draw (7) coordinate[c3,fill=brown];

\draw (23) coordinate[c2,fill=green];
\draw (23) coordinate[c3,fill=Emerald];

\draw (29) coordinate[c2,fill=Tan];

\draw (21) coordinate[c2,fill=DarkOrchid];

\draw (3) coordinate[c2,fill=blue];
\draw (3) coordinate[c3,fill=YellowGreen];

\draw (11) coordinate[c2,fill=SpringGreen];

\draw (9) coordinate[c2,fill=Apricot];
\draw (9) coordinate[c3,fill=Mulberry];

\draw (27) coordinate[c2,fill=Emerald];
\draw (27) coordinate[c3,fill=black];

\draw (13) coordinate[c2,fill=Tan];
\draw (13) coordinate[c3,fill=Fuchsia];

\draw (7) coordinate[c2,BrickRed];
\draw (7) coordinate[c3,fill=brown];

\draw (8) coordinate[c2,fill=pink];

\draw (19) coordinate[c2,fill=CornflowerBlue];

\draw (16) coordinate[c2,fill=CornflowerBlue];
\draw (16) coordinate[c3,fill=pink];

\draw (1) coordinate[c2,fill=Salmon];
\draw (1) coordinate[c3,fill=Fuchsia];

\draw (17) coordinate[c2,fill=PineGreen];
\draw (17) coordinate[c3,fill=DarkOrchid];

\draw (15) coordinate[c2,fill=SpringGreen];
\draw (15) coordinate[c3,fill=Salmon];

\draw (25) coordinate[c2,fill=Mulberry];
\draw (25) coordinate[c3,fill=black];

\draw (22) coordinate[c2,fill=YellowGreen];
\draw (22) coordinate[c3,fill=MidnightBlue];

\draw (28) coordinate[c2,fill=RawSienna];

\draw (10) coordinate[c2,fill=BlueGreen];

\draw (6) coordinate[c2,fill=PineGreen];

\draw (2) coordinate[c2,fill=gray];

\end{tikzpicture}
\end{center}
\caption{Greechie orthogonality diagram of a logic~\cite[Fig.~2, p.~102201-8]{2015-AnalyticKS}
realizable in $\mathbb{R}^3$
with the true--implies--value indefiniteness (neither true nor false) property on the atoms $\vert {\bf a} \rangle $ and $\vert {\bf b} \rangle $,
respectively.
The 8 classical value assignments require $\vert {\bf a} \rangle $ to be false.
Therefore, if  one prepares the quantized system in state $\vert {\bf a} \rangle $,
the second state $\vert {\bf b} \rangle $ cannot have any consistent classical value assignment -- it must be value indeterminate/indefinite.
}
\label{2019-s}
\end{figure}
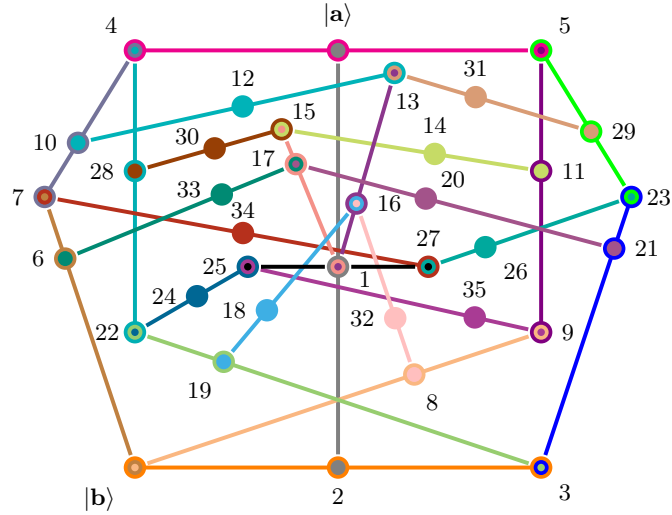

Let me, for the sake of presenting another quantum resource, contemplate one example for which, relative to the assumptions made, quantum ``computation'' outperforms
classical recursion theory: the generation of (allegedly) irreducibly indeterministic numbers; or sequences thereof~\cite{2014-nobit}.
A recent extension of the
Kochen-Specker theorem~\cite{2012-incomput-proofsCJ,PhysRevA.89.032109,2015-AnalyticKS}
allowing partial value assignments suggests the following algorithm:
Suppose one prepares a quantized system capable of three or more mutually exclusive outcomes,
formalized by Hilbert spaces of dimension three and higher, in an arbitrary pure state.
Then,
relative to certain reasonable assumptions (for value assignments and noncontextuality),
this system cannot be in any defined, determined property in any other direction of Hilbert state not
collinear or orthogonal to the vector associated with the prepared state~\cite{pitowsky:218,hru-pit-2003}:
the associated classical truth assignment cannot be a total function.
The proof by contradiction is constructive and involves a configuration of intertwining quantum contexts (aka orthonormal bases).
Figure~\ref{2019-s} depicts a particular configuration of quantum observables, as well as a particular one of their faithful orthogonal representations~\cite{lovasz-79}
in which the prepared and measured states are an angle
$
\text{arccos } \langle {\bf a}\vert {\bf b} \rangle =
\text{arccos} \left[
\begin{pmatrix}
1,0,0
\end{pmatrix}   \frac{1}{2}
\begin{pmatrix}
\sqrt{2},1,1
\end{pmatrix}^\intercal
\right] = \frac{\pi}{4}$ apart~\cite[Table~1]{2015-AnalyticKS}.
\newif\iflabel \labeltrue

Whenever one approaches quantum indeterminacy from the empirical, inductive side,
one has to recognize that, without {\it a priori} assumptions, formal proofs of (in)computability,
and more so algorithmic incompressibility (aka randomness~\cite{ml:70})
are blocked by reduction to the halting problems and similar~\cite{svozil-2016-pu-book}.
The best one can do is to run
tests, such as Borel normality and other criteria, on finite sequences of random number generators~\cite{PhysRevA.82.022102,Abbott_2019}
which turn out to be consistent with the aforementioned value indefiniteness and
quantum indeterminacy.

\section{Afterthoughts on assumptions}

Let me, as a substitute for a final discussion, mention a {\it caveat}: as all results and certifications hold true relative to the assumptions made,
different assumptions and axioms may change the perceptual framework and results entirely.
One might, for instance, disapprove of the physical existence of states and observables beyond a single vector or context~\cite{svozil-2018-whycontexts,Auffeves-Grangier-2018}.
Thereby, the problem of measuring other contexts would be relegated to the general measurement problem of coherent superpositions~\cite{london-Bauer-1983}.
In this case, as von Neumann, Wigner and Everett have pointed out, by ``nesting'' the measurement objects and the measurement apparatus in larger and larger systems~\cite{everett-collw},
the assumption of universal validity of the quantum state evolution would
result in a mere epistemic randomness; very much like the randomness encountered in, and the second law of~\cite{Myrvold2011237}, classical statistical physics.
From that perspective quantum randomness might turn out to be valid ``for all practical purposes''~\cite{bell-a} through interaction with a huge number of (uncontrollable) degrees of freedom
in the environment of a quantized system in a coherent state, ``squeezing'' out this coherence very much like a balloon losing gas~\cite{Zyczkowski-balloon}.

\section*{Acknowledgments}
I kindly acknowledge enlightening discussions with Cristian Calude about many of the subjects mentioned.
All misconceptions and errors are mine.


%
%
%
%

\end{document}